# Magnetic anisotropy and lattice dynamics in FeAs studied by Mössbauer spectroscopy

A. Błachowski[1], K. Ruebenbauer[1*], J. Żukrowski[2], and Z. Bukowski[3]

[1]Mössbauer Spectroscopy Division, Institute of Physics, Pedagogical University
ul. Podchorążych 2, PL-30-084 Kraków, Poland

[2]AGH University of Science and Technology, Faculty of Physics and Applied Computer Science, Department of Solid State Physics
Av. A. Mickiewicza 30, PL-30-059 Kraków, Poland

[3]Institute of Low Temperature and Structure Research, Polish Academy of Sciences
ul. Okólna 2, PL-50-422 Wrocław, Poland

[*]Corresponding author: sfrueben@cyf-kr.edu.pl

### Abstract

Iron mono-arsenide in the powder form has been investigated by transmission $^{57}$Fe Mössbauer spectroscopy in the temperature range 4.2 – 1000 K. Additional spectra have been obtained at 20 K and 100 K applying external magnetic field of 7 T. It was found that the spin spiral propagating along the *c*-axis leads to the complex variation of the hyperfine magnetic field amplitude with the spin orientation varying in the *a-b* plane. The magnitude of the hyperfine field pointing in the direction of the local magnetic moment depends on the orientation of this moment in the *a-b* plane. Patterns are vastly different for iron located in the $[0\,k\,0]$ positions and for iron in the $[0\,k+\tfrac{1}{2}\,0]$ positions within the orthorhombic cell set to the *Pnma* symmetry. Lattice softens upon transition to the paramagnetic state at 69.2 K primarily in the *a-c* plane as seen by iron atoms. This effect is quite large considering lack of the structural transition. Two previously mentioned iron sites are discernible in the paramagnetic region till 300 K by different electron densities on the iron nuclei. The anisotropy of the iron vibrations developed at the transition to the paramagnetic state increases with the temperature in accordance with the harmonic approximation, albeit tends to saturation at high temperatures indicating gradual onset of the *quasi*-harmonic conditions. It seems that neither hyperfine fields nor magnetic moments are correct order parameters in light of the determined static critical exponents. Sample starts to loose arsenic at about 1000 K and under vacuum.

## 1. Introduction

The iron mono-arsenide FeAs seems very interesting compound despite simple chemical formula as there is a competition between highly directional covalent Fe-As-Fe and metallic bonds due to the fact that the compound is metallic with complex narrow-band structure. FeAs crystallizes in the orthorhombic structure as determined by Hägg [1], resolved by Fylking [2] and with details determined by Selte and Kjekshus [3]. They found $a$=5.4420 Å, $b$=3.3727 Å and $c$=6.0278 Å at room temperature [3]. The compound is formed within extremely narrow composition range with the very high atomic order [3] confirming importance of the covalent bonds. There are four formal FeAs molecules per chemical unit cell with each iron atom being coordinated by six arsenic atoms forming distorted octahedron and vice versa. The structure could be refined in the $Pnma$ space group, albeit somewhat better refinements were obtained for the lower symmetry $Pna2_1$. It was found that thermal ellipsoids are almost isotropic for arsenic atoms, while for the iron atoms one has almost axially symmetric ellipsoid with the longest axis aligned with the $b$-axis. The $[0\frac{1}{2}0]$ plane is a mirror plane for the $Pnma$ symmetry, while for the $Pna2_1$ symmetry either two arsenic and/or two iron atoms are moved along the $b$-axis from the regular positions in several possible ways leading to two inequivalent pairs of iron sites in the cell. For the $Pnma$ symmetry all iron atoms are equivalent with the arsenic octahedron characterized by single shortest and longest Fe-As distances and two pairs of remaining distances with distances being equal each other within the pair. In the case of the $Pna2_1$ symmetry octahedrons around iron atoms lying close to the mirror plane of the $Pnma$ symmetry transform into octahedrons with all distances being different one from another, albeit both of these octahedrons are mutually equivalent. Similar results were obtained by Lyman and Prewitt [4] with the assumption that all thermal ellipsoids are isotropic. They investigated FeAs at room temperature versus pressure up to 8.3 GPa without finding any structural phase transition. Crystal structure of FeAs is shown in Figure 1. Detailed X-ray, neutron diffraction and magnetometric studies were performed versus temperature by Selte *et al.* [5]. The anomalous thermal expansion was found for the $a$-axis up to about 600 K (see also [6]). Magnetic susceptibility deviates from the Curie-Weiss law below 300 K and above 650 K. The paramagnetic moment was found to be about 3.1 $\mu_B$ per iron atom. FeAs orders magnetically at ~70 K [7] with the magnetic moment per iron atom being about 0.5 $\mu_B$ at 12 K. A transition to the high spin state with the increasing temperature is confirmed by the heat capacity measurements performed till above the melting point of about 1325 K [8]. The excess specific heat is due to the significant re-population of the narrow 3d bands carrying electron spins. Double antiferromagnetic (interlaced [7]) spiral is formed with the moments lying in the $a$-$b$ plane, and with the propagation vector aligned with the $c$-axis. The propagation vector is incommensurate with the $c$-axis amounting to $0.375 \times (2\pi)/c$ at 12 K [5] or $0.395 \times (2\pi)/c$ at 4 K [7]. Polarized neutron scattering revealed that spirals have elliptic deformation with the longest axis of the ellipse being aligned with the $b$-axis [7]. It was found that the moment along the $b$-axis is about 15 % larger than along the $a$-axis [7]. There are two spirals related one to another by the relative phase angle [5, 7]. One of them is passing through the iron atoms lying in the mirror plane of the $Pnma$ space group [7]. According to the results obtained by Rodriguez *et al.* [7] there are altogether four spirals per chemical unit cell passing through the iron atoms projected on the $a$-$b$ plane within the unit cell with the constant angle increment while moving by two lattice constants $c$ along the $c$-axis (antiferromagnetic order). The angle increment is defined in the $a$-$b$ plane around the $c$-axis. Due to the fact, that the spiral period is incommensurate with the lattice constant $c$ one obtains practically *quasi-*continuous distribution of moments on iron atoms and within the $a$-$b$ plane. For elliptic spiral magnetic moment (absolute value) depends on the above angle counted let us say from the $a$- axis in the right hand convention. The structure and transport properties have been investigated versus temperature and pressure by Jeffries *et al.* [9]. No structural phase transitions were found, albeit the magnetic moment was found to vanish at about 11 GPa. A reentrant behavior of the Hall coefficient found by Segawa and Ando [10] indicates complex band structure. Muon precession studies revealed presence of two stopping sites with different local magnetic fields [11]. Early theoretical studies of the spiral magnetic structure in FeAs were performed by Morifuji and Motizuki [12]. Dobysheva and Arzhnikov [13] performed calculations of the electronic structure within density functional theorem formalism obtaining results in good agreement with the available experimental data including spiral magnetic structure. Calculations were performed for the *Pnma* setting. Similar calculations were performed by Parker and Mazin [14]. Complex band structure has been confirmed with the density of states on the Fermi surface being quite sensitive to the kind of the magnetic order [14]. No nests within the Fermi surface were found excluding coplanar spin density wave (SDW) type of magnetism [14]. Early 14.41-keV $^{57}$Fe transmission Mössbauer spectra were obtained in the non-magnetic region [15-17]. Spectra versus temperature were measured by Kulshreshtha and Raj [18] including magnetic region (see also [19]). High temperature spectra are slightly broadened doublets with the left (lower velocity) line being systematically more intense than the right line. Spectra in the magnetically ordered region were found to be very complex [18]. Subsequent work by Häggström *et al.* [20] shed more light on the problem. The high temperature spectra were found similar to the previous ones. For the low temperature spectra it was found that the electric field gradient tensor (EFG) has positive principal component lying in the *a-c* plane and making 34° angle with the *c*-axis. The asymmetry parameter was found to be unity with the second negative component of the EFG aligned with the *b*-axis. Due to the extreme value of the asymmetry parameter some other equivalent orientations of the EFG could not be excluded [20]. The hyperfine magnetic field was distributed in the *a-b* plane. A reasonable fit was obtained by using five sites of unequal populations with different fields distributed between *a* and *b* axes. The contour made by the hyperfine field had distortion far beyond the elliptic distortion indicating lack of proportionality between the hyperfine field and the magnetic moment on the iron atom and need to include higher order terms than the second order terms in description of the hyperfine field distribution [20]. No correlation between particular hyperfine field and either EFG or total shift was found confirming metallic environment on the local scale [20].

Discovery of the iron-based superconductors containing Fe-As layers makes investigation of the Fe-As bonds quite important [21]. Superconductors have tetragonal coordination of iron by arsenic and they are definitely more layered than FeAs. They are more susceptible to substitution by foreign atoms as well.

Hence, it seems desirable to reinvestigate FeAs by means of the Mössbauer spectroscopy in order to answer the following questions. How many iron-sites are distinguishable in the magnetic and non-magnetic regions from the point of view of EFG and total (isomer) shift? Is any dynamic anisotropy (temperature dependent) on the iron sites, i.e., anisotropy of the recoilless fraction? One has to note that diffraction methods mix anisotropy of the thermal motion and small static disorder, the latter being susceptible to the variation of the temperature in some cases. On the other hand, local methods like the Mössbauer spectroscopy mix thermal anisotropy with the preferential orientation (texture) in the powder samples. Such orientation does not depend on the temperature for a given sample. The next question is. How many magnetic sites are present and what is the shape of the hyperfine fields versus rotation in the *a-b* plane? The unusual static critical exponent 0.16 concerned with the magnetic order requires some closer look as well [7].

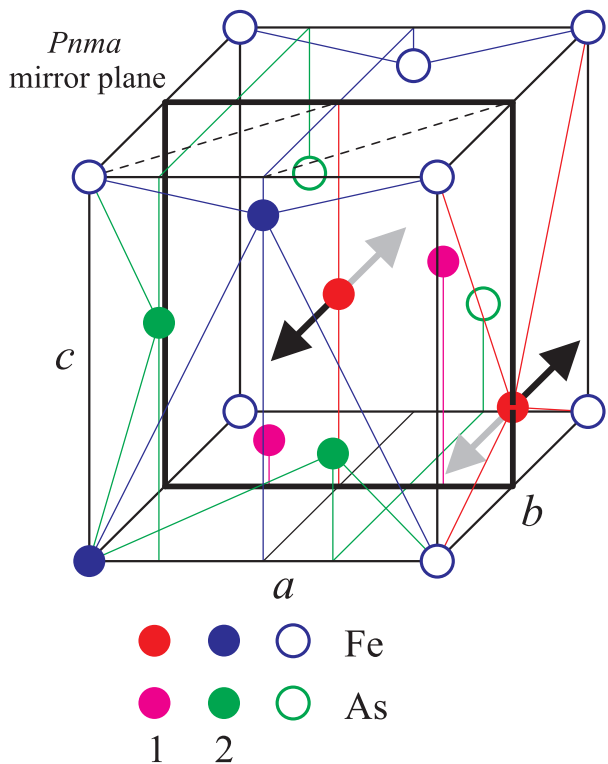

**Figure 1** Crystal structure of FeAs. Empty circles denote atoms belonging to the adjacent chemical cells. Arrows aligned with the *b*-axis show one of possible distortions (most likely) from the *Pnma* symmetry. Distortions shown by black and gray arrows seem equally probable and they can interchange between themselves on the quite local scale. Iron atoms marked in red (1) and blue (2) are equivalent without distortion. Iron 1 is located at or near $[0\, k+\tfrac{1}{2}\, 0]$ planes, while iron 2 is located at $[0\, k\, 0]$ planes with $k$ being the Miller index.

## 2. Experimental

FeAs in a single-crystalline form (small crystals having largest dimension of ~2 mm) was prepared using Sn flux technique. FeAs powder and Sn pieces taken in the molar ratio FeAs:Sn = 5:95 were loaded into an alumina crucible and sealed in an evacuated silica tube. The ampoule was heated up to 950º C, kept at this temperature for 10 hours and slowly (2º C/h) cooled down to 500º C. Next, the ampoule was turned over and liquid Sn flux was decanted. Finally, FeAs single crystals were etched in diluted hydrochloric acid in order to remove residues of tin. The composition of the as grown crystals was confirmed by energy dispersive X-ray analysis (EDX) and the crystal structure was checked by using powder X-ray diffraction.

The absorber for Mössbauer spectroscopy measurements was made mixing 57 mg of FeAs crushed to the fine powder with the ample amount of $B_4C$ fine powder. Absorber for the null external magnetic field measurements was sandwiched between two both-side aluminized mylar windows of 0.1 mm thickness each. Windows were attached to respective parts of the copper sample holder. The total diameter of the circular absorber amounted to 16 mm with the clear bore of 12 mm. Hence, one had 28.3 mg/cm² of FeAs. The same sample (powder mixture) was repacked for measurements in the external magnetic field into similar holder having 18 mm diameter with the effective clear bore of about 10 mm. Hence, the absorber thickness changed to 22.4 mg/cm² of FeAs. A commercial $^{57}$Co(Rh) source delivered by Ritverc G.m.b.H. having active diameter 8 mm and of activity 50 mCi was used. The source was made movable and it was maintained at ambient conditions in all measurements (~24 °C). Horizontal transmission geometry was adopted and resonant photons were counted by using Kr-filled proportional counter supplied by LND Inc. Measurements in the null external field were performed by means of the MsAa-3 spectrometer supplied by RENON and registering photons in the 14.41-keV line photo-peak. Spectra were measured for the absorber maintained at temperatures between 4.2 K and 300 K with the stability better than 0.01 K. The liquid helium SVT-400TM cryostat made by Janis Research Co. was used to maintain the absorber temperature. Spectra in the external field of 7 T aligned anti-parallel with the beam propagation direction were obtained using 7TL-SOM2-12 MOSS liquid helium cryostat of Janis Research Co. equipped with the superconducting magnet. Temperature stability was comparable to the previous case. Detector of the same type as previous and source were shielded from the stray field. Spectra were collected by means of the MsAa-4 spectrometer delivered by RENON. Spectra obtained in the photo-peak and Kr K-escape peak were added to improve statistics. Background under lines used to collect the signal remained practically constant for each series of measurements and detector worked within the reasonable linear range considering amplitude and frequency response. Velocity scales of both spectrometers were calibrated by the Michelson-Morley interferometers equipped with the metrologic quality He-Ne lasers. The velocity scale was corrected for the γ-ray beam divergence, if required. A round-corner triangular reference function was used for both spectrometers. The source average position was kept constant and the frequency of the transducer was selected in such way to minimize amplitude of the source motion, while staying in the best frequency response range of the transducer. Spectra above 300 K were obtained applying water-cooled vacuum transmission oven with the boron nitride sample holder and under dynamic vacuum of about $10^{-6}$ hPa. The same powder mixture as used for previously performed measurements was applied. Temperature stability was about 0.1 K. Spectra were calibrated and folded with the estimation of the background in the γ-ray direct spectra windows used to collect data by means of the Mosgraf-2009 suite proper applications [22]. All spectral shifts are reported versus room temperature α-Fe.

3. Data treatment

Mössbauer spectra obtained in the magnetically ordered state (in the null external magnetic field and ambient pressure) were fitted using GMFeAs application of the Mosgraf-2009 suite [22]. A standard transmission integral was applied for the single line unpolarized and resonantly thin source in the narrow geometry approximation. The (random) absorber was considered as unpolarizing and at temperature high enough to equalize occupation of the ground hyperfine levels. Two iron sites with equal effective populations were necessary to fit the data in the semi-classical approximation. The isotropic absorber conditions are satisfied in this temperature range. It was found that EFG is aligned with the main crystal axes and it exhibits very large anisotropy parameter common for both sites $\eta = (V_{11} - V_{22})/V_{33}$ with $|V_{11}| \leq |V_{22}| \leq |V_{33}|$ denoting principal components of EFG. Note that EFG is traceless tensor. On the other hand, the value of the positive quadrupole coupling constant varies between sites as well as the total shift. One has to note that the electric quadrupole interaction is seen for the 14.41-keV transition in $^{57}$Fe solely in the (first) excited nuclear state as the ground nuclear state spin equals $I_g^{(\pi_g)} = \frac{1}{2}^{(-)}$, while the excited state spin equals $I_e^{(\pi_e)} = \frac{3}{2}^{(-)}$. A single photon transition is almost pure M1 transition. The symbols $\pi_e$ and $\pi_g$ denote parities of the respective nuclear states. The spectroscopic nuclear quadrupole moment of the excited state amounts to $Q_e = +0.17$ b [23], and hence one can conclude that $V_{33}$ component of the EFG is positive. The quadrupole coupling constant is expressed here as $A_Q = \dfrac{eQ_e V_{33}}{4I_e(2I_e - 1)}\left(\dfrac{c}{E_0}\right)$. The symbol e stands for the positive elementary charge. The symbol $c$ denotes speed of light in vacuum, while the symbol $E_0 > 0$ stands for the nuclear transition energy. The absorber line-width is common for both sites and lines exhibit almost natural width. The hyperfine magnetic field was found to be restricted to the $a$-$b$ plane for both sites. The situation is sketched in Figure 2.

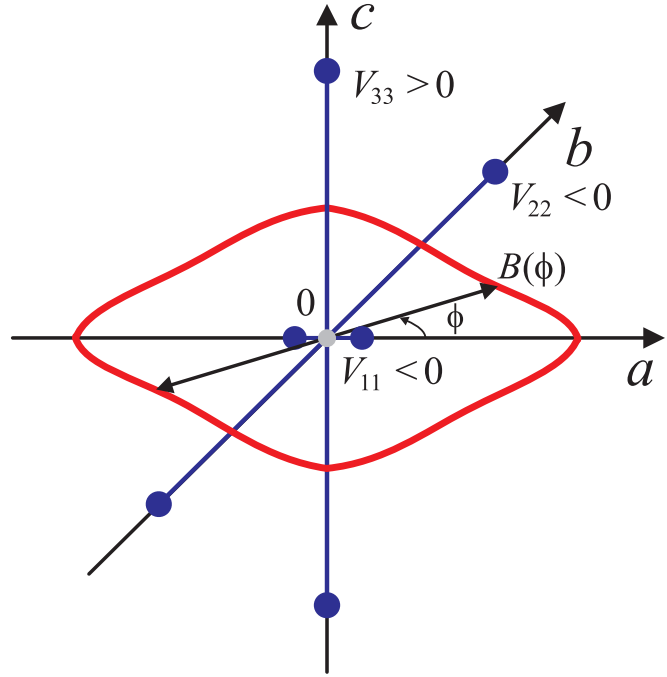

**Figure 2** Orientation of the EFG and hyperfine magnetic field in the main crystal axes for both iron sites.

Magnetic hyperfine field $B(\phi)$ depends on the angle $\phi$ measured by the right hand rotation from the $a$-axis in the $a$-$b$ plane. The same convention is used for both sites. Above field has been parameterized by the following expression independently for each site:

$$B(\phi) = B_0 \exp\left[\sum_{l=1}^{L}\sum_{m=0}^{l}\left(\dfrac{l!}{(l-m)!m!}\right) P_{lm} \cos^{l-m}(\phi) \sin^{m}(\phi)\right].$$

(1)

The factor $B_0 > 0$ stands for the scaling field, while the real coefficients $P_{lm}$ are components of the fully symmetric real tensors $\mathbf{P}_l$ of the order $l$ and operating in two spatial dimensions. They have $l+1$ irreducible components for the order $l$. The maximum order has been restricted to $L = 4$. Odd tensors with $l = 1, 3$ were dropped due to the satisfaction of the antiferromagnetic conditions. The following non-zero coefficients were essential to fit data at low temperature: $P_{21}, P_{22}, P_{41}, P_{42}$. For each site 512 equally spaced in the range $(0 : 2\pi)$ configurations were used and for each configuration complete diagonalization of the hyperfine Hamiltonians was applied.

Spectra obtained in the same spectrometer setting, albeit above the magnetic transition were fitted applying GMFP application of the Mosgraf-2009 suite [22] within standard transmission integral approximation as well. Complete diagonalization of the hyperfine Hamiltonians was used as above. Two sites of equal effective population are still present. They could be fitted with very similar quadrupole splittings, common line-width (somewhat broader than in the magnetic region), but different spectral shifts (Model 1). The fitting model with almost the same shifts, but different splittings works equally well (Model 2). Some anisotropy of the line intensities with the lower line being more intense develops above transition to the disordered magnetic state. The quadrupole interaction was treated in this region by using common for both sites asymmetry parameter fixed on the value obtained at highest temperature of the magnetic region. The quadrupole coupling constants were fitted independently for each site as well as the total shifts. It was assumed that a thermal ellipsoid of the axial symmetry and common for both sites develops with the short axis aligned with the *b*-axis. The choice of the common ellipsoid orientation and internal symmetry followed Ref. [3]. Such ellipsoid requires the smallest deformation from the spherical symmetry within all possible ellipsoids aligned with the quantization frame *abc* to account for the observed spectrum anisotropy. For such case relative intensities of each spectral doublet are perturbed due to the introduction of the following operator to the absorption cross-section $\langle \lambda_{eg} | M \rangle \langle M | \mathbf{C} | M' \rangle \langle M' | \lambda_{eg} \rangle$. Here the symbol $\lambda_{eg}$ denotes transition energy between ground hyperfine level $g$ and the excited hyperfine level $e$. Symbols $M = 0, \pm 1$ and $M' = 0, \pm 1$ stand for the magnetic quantum numbers of the absorbed radiation, while the Hermitean operator $\mathbf{C}$ differs from the unit operator for the anisotropy of the recoilless fraction and/or preferential orientation within the unpolarizing absorber. In general, this operator takes on the following form for dipolar transitions:

$$\mathbf{C} = \begin{pmatrix} g_{11} & -g_{10}^* & g_{1-1}^* \\ -g_{10} & 1 & g_{10}^* \\ g_{1-1} & g_{10} & g_{11} \end{pmatrix}.$$

(2)

Coefficients $g_{MM'}$ could be calculated for the homogeneous absorber (on all relevant scales) as $g_{MM'} = \alpha_{MM'} / \alpha_{00}$. On the other hand, coefficients $\alpha_{MM'}$ follow the relationship [24, 25]:

$$\alpha_{MM'} = \int_0^{2\pi} d\gamma \exp[i(M - M')\gamma] \int_0^{\pi} d\beta \sin(\beta) f(\beta\gamma) \sum_{k=\pm 1} d_{kM}^{(1)}(\beta) d_{kM'}^{(1)}(\beta).$$

(3)

The function $f(\beta\gamma) > 0$ stands either for the recoilless fraction and/or for the probability to find the orientation $\beta\gamma$ in the quantization coordinates used (here *abc*) with the angles $\beta$ and $\gamma$ denoting polar and azimuthal angle, respectively. The symbol $d_{kM}^{(1)}(\beta)$ denotes generalized spherical harmonic. For a thermal ellipsoid mentioned above one has $f(\beta\gamma) \sim \exp[q^2 \Delta b \sin^2 \beta \sin^2 \gamma]$. Here the symbol $q = E_0 /(\hbar c)$ stands for the wave number of absorbed radiation with $\hbar$ denoting Planck constant divided by $2\pi$. The symbol $\Delta b = b_{33} - b_{22} = b_{11} - b_{22} > 0$ describes thermal ellipsoid anisotropy in the reference frame *abc*. For such setting one obtains as non-zero the following coefficients $\frac{1}{2} < g_{11} < 1$ and $-\frac{1}{2} < \text{Re}(g_{1-1}) < 0$. One has to note that for $g_{11} < \frac{3}{2}$ the following relationship is satisfied quite accurately $g_{11} - \text{Re}(g_{1-1}) \approx 1$. Hence, the parameter $\text{Re}(g_{1-1})$ has been fitted to the data and the relationship $g_{11} - \text{Re}(g_{1-1}) = 1$ has been applied. The spectrum is sensitive to the parameter $\text{Re}(g_{1-1})$ due to the fact that EFG has no axial symmetry. It has to be noted that only single real parameter describing relative intensities within spectral doublet could be fitted. Strong temperature dependence of $\Delta b$ is an indication that the anisotropy has dynamical character and the absorber is almost randomly oriented. The same fitting models were used for spectra obtained above 300 K.

Spectrum obtained in the homogeneous external magnetic field oriented anti-parallel to the beam propagation direction and above magnetic transition was fitted by means of the GMFPILVD application of the Mosgraf-2009 suite [22]. The source was shielded from the stray field and hence retained previously described properties. A standard double-transmission integral approximation was used in order to account for the polarizing properties of the absorber in the external field [26]. An obvious assumption was made that both polarization states are equally populated in the beam leaving the source. Due to the fact that absorber had almost random orientation the cross-section was averaged over 16 x 32 equally spaced points over the unit sphere (16 different polar angles and 32 different azimuthal angles) with respect to the external field. Due to the alignment of the beam with the field there is no need to average over the third Eulerian angle. The local anisotropy of the recoilless fraction was accounted for by using $q^2 \Delta b$ parameter described above. The value was taken from data obtained in the null field.

## 4. Results

Figure 3 shows selected spectra obtained in the temperature range 4.2 – 300 K in the null external field. Spectra within the range 72 – 300 K do not exhibit magnetic order. The line-width amounts to about 0.14 mm/s at and above 74 K, while it equals 0.16 mm/s at 72 K indicating onset of the magnetic order. Spectra within the range 4.2 – 67 K exhibit fully developed magnetic spirals. Parameters were obtained for the spectrum at 4.2 K and used as starting parameters for subsequently higher temperatures. The asymmetry parameter equals $\eta = 0.88(1)$ at 4.2 K and does not change at higher temperatures. Line-width remained in the range 0.11 – 0.14 mm/s with the increasing temperature, staying at lower value till about 60 K. Parameters of the magnetic (spin) spirals are gathered in Table I for selected spectra. Simplified spiral without parameters $P_{21}$ and $P_{42}$ was used in the range 68 K – 70 K with the line-width increasing till 0.16 mm/s.

Figure 4 shows both magnetic spirals for selected temperatures. The magnetic hyperfine field is restricted to the *a-b* plane. In reality *a*-axis and *b*-axis shown in Figure 4 have no arrows, as EFG is invariant against axis inversion. Hence, one has four equivalent representations (in definite plane – eight in the three dimensional space) reducing to two in the case of inversion center being present due to antiferromagnetism. This statement applies to both sites of iron – on the local scale. More asymmetric spiral (right column) is likely to belong to the perfectly ordered sites, i.e., to the iron in [0 *k* 0] planes. Regions with the larger hyperfine field follow approximately direction of the Fe-As bonds projected on the *a-b* plane for the more asymmetric spiral. Hence, it seems that the electron spin polarization is distributed among anisotropic conduction bands and more localized 3d electrons strongly mixed with the p electrons of arsenic. Subsequent iron core polarization (Fermi field) follows above spin polarization. More featureless spiral (left column) is probably more symmetric due to the larger contribution of the itinerant spins and due to some disorder along the *c*-axis, and hence it is likely to originate on the $[0\,k+\tfrac{1}{2}\,0]$ iron atoms. Dissimilarity between magnetic moment and the hyperfine field of the atom with above moment is clearly seen here [27]. Hence, the magnetic hyperfine field on the iron nucleus could be expressed as $B(\phi) = A(\phi)\mu(\phi)$ with $\mu(\phi) = \mu_0(1+0.15\sin^2\phi)$ according to Ref. [7]. Here the symbol $\mu_0$ denotes the value of the magnetic moment along the *a*-axis. The absolute value of magnetic hyperfine coupling constant $A(\phi)$ [27] varies with the angle $\phi$ in different fashion for two groups of sites.

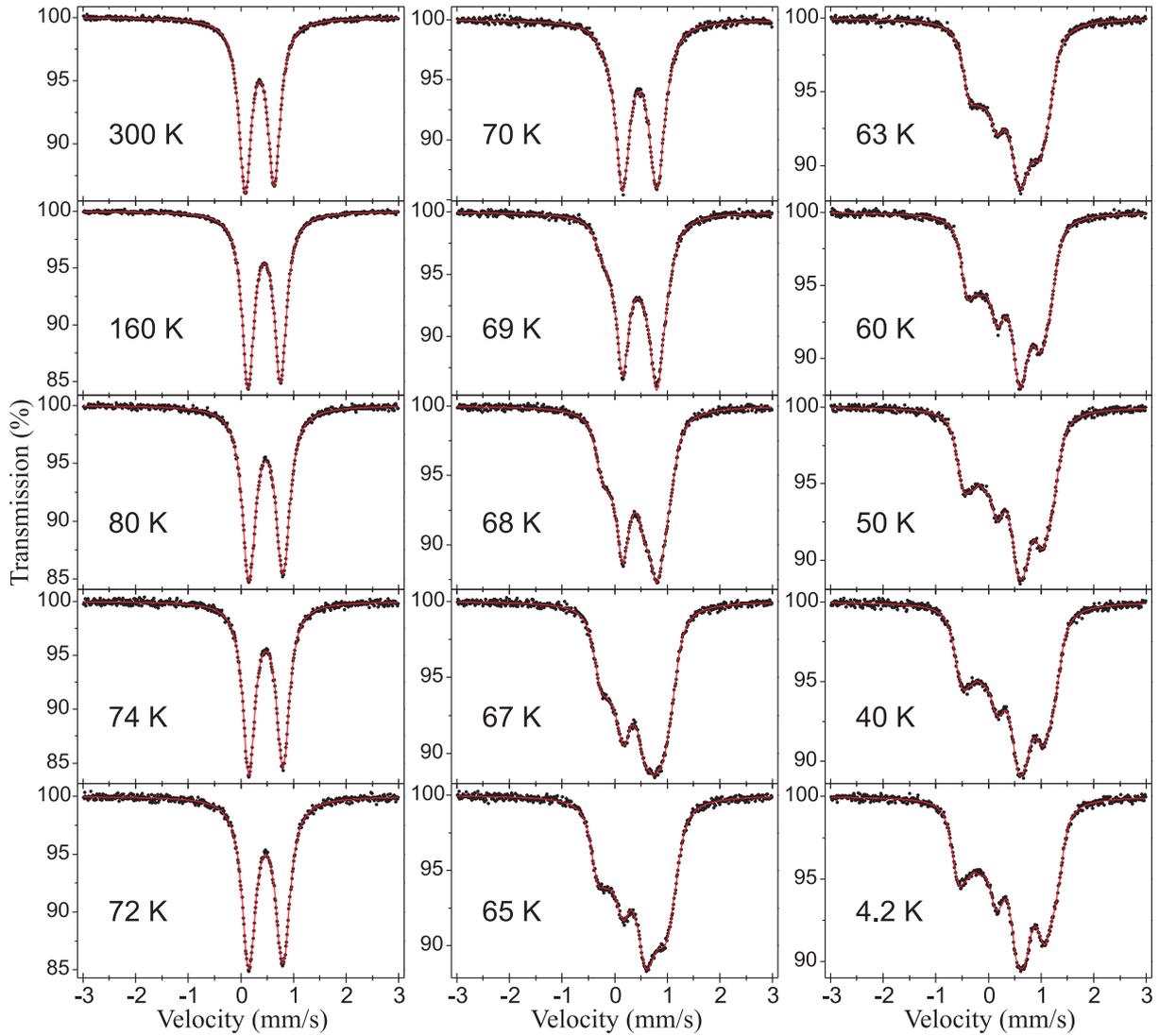

**Figure 3** $^{57}$Fe Mössbauer spectra of FeAs obtained within temperature range 4.2 – 300 K in the null external magnetic field.

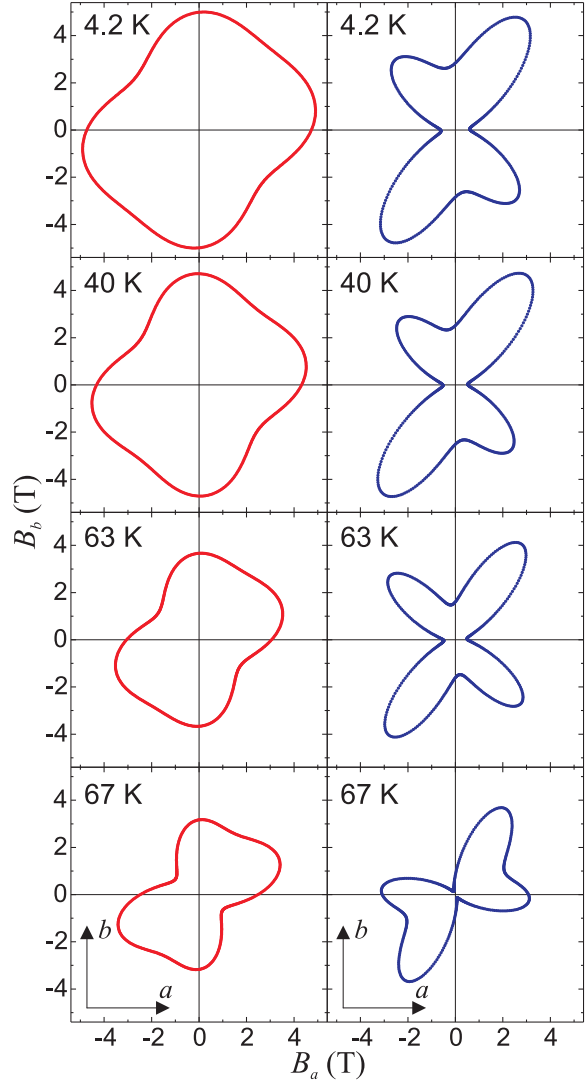

**Figure 4** Anisotropy of the hyperfine magnetic fields in FeAs for selected temperatures. Left column shows $[0\,k+\tfrac{1}{2}\,0]$ iron, while the right column shows $[0\,k\,0]$ iron. The symbol $B_a$ denotes hyperfine field component on iron along the $a$-axis, while the symbol $B_b$ stands for the iron hyperfine field component along the $b$-axis.

**Table I**

Parameters $B_0$ and $P_{lm}$ used to fit magnetic spirals at temperatures of Figure 4. Upper rows for each temperature correspond to $[0\,k+\tfrac{1}{2}\,0]$ iron, while the lower rows to $[0\,k\,0]$ iron.

| $T$ (K) | $B_0$ (T) | $P_{21}$ | $P_{22}$ | $P_{41}$ | $P_{42}$ |
|---|---|---|---|---|---|
| 4.2 | 4.73(5) | 0.05(3) | 0.05(1) | 0.08(2) | -0.13(1) |
|  | 0.62(3) | 0.5(1) | 1.52(4) | -0.4(1) | 0.73(5) |
| 40 | 4.3(1) | 0.0(1) | 0.09(2) | 0.13(4) | -0.15(2) |
|  | 0.53(3) | 0.5(2) | 1.56(5) | -0.4(2) | 0.81(5) |
| 63 | 3.1(1) | 0.0(1) | 0.18(4) | 0.2(1) | -0.12(4) |
|  | 0.5(1) | 0.6(4) | 1.1(1) | -0.6(5) | 1.0(1) |
| 67 | 2.5(3) | 0.1(1) | 0.2(1) | 0.4(1) | -0.2(1) |
|  | 3.1(2) | 2.6(1) | -1.3(3) | -1.4(2) | -0.3(2) |

Average hyperfine magnetic fields for both iron sites are plotted versus temperature in Figure 5. They were fitted using the same method as described in detail in Ref. [28]. The parameter $T_c$ stands for the transition temperature, while the parameter $\alpha$ denotes static critical exponent. A small difference in $T_c$ for two fields is likely to be due to the fact that the smaller field exhibits much sharper decay. Exponent for the larger field is similar to the exponent obtained by Rodriguez et al. [7] as well as transition temperature. On the other hand, exponent for the smaller field is exceptionally small like for the first order non-magnetic transition (like metal-insulator) with the magnetic moment on the one side only. Such situation occurs for the first order transition in MnAs [29, 30]. However, neither latent heat nor hysteresis is observed in the present case of FeAs [8]. Crystal structure does not change at the magnetic ordering, too [7]. Therefore it seems that neither magnetic moments nor hyperfine fields are correct order parameters for FeAs and the magnetic transition is driven by some underlying transition of different kind e.g. some order-disorder transition with negligible latent heat and hysteresis. One can imagine ordering of the iron displacements from the $[0\,k+\tfrac{1}{2}\,0]$ planes. Such ordering would cause line narrowing observed at low temperature, indeed. Very narrow temperature range of the incoherent magnetic order [28] supports hypothesis of the underlying transition leading to the magnetic order.

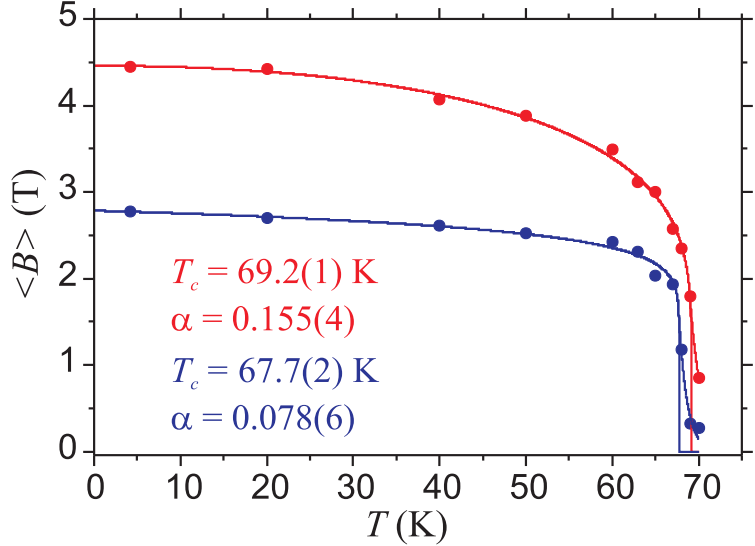

**Figure 5** Average hyperfine magnetic fields $\langle B \rangle$ plotted versus temperature. Larger values correspond to the $[0\,k+\tfrac{1}{2}\,0]$ iron, while the smaller values belong to $[0\,k\,0]$ iron. The symbol $T_c$ stands for the transition temperature, while the symbol $\alpha$ stands for the static critical exponent. The upper set of values describes larger field, while the lower set of values smaller field. For details see text and Ref. [28].

Figure 6 shows spectral shift $S$ and quadrupole coupling constant $A_Q$ versus temperature. For non-magnetic spectra two equivalent models described above could be fitted. Model 1 with very similar quadrupole splittings, common line-width, but different spectral shifts and Model 2 with almost the same shifts, but different splittings. Model 1 seems more realistic as the quadrupole coupling constants should be very similar for both iron sites due to the very small departure from the *Pnma* symmetry. In the magnetic region there is no distinction between models as solutions are unique. Quadrupole coupling constant changes at the onset of magnetism due to the magnetostriction and following redistribution of 3d electrons. The effect is much more pronounced for the almost perfectly ordered iron in [0 k 0] planes, where the quadrupole coupling constant diminishes upon transition to the magnetically ordered state. Decrease of the quadrupole coupling constant with increasing temperature above magnetic transition is consistent with earlier results (see [18]). Shifts are harder to interpret at the onset of the magnetic order as the lattice becomes more rigid in the magnetic state - see below. The electron density on the iron nucleus diminishes for $[0\,k+\tfrac{1}{2}\,0]$ iron, while for the remaining iron one cannot get sure conclusion due to the competition between jump in the second order Doppler shift (SOD) and jump of the electron density. However, the electron density difference between two sites amounts to 0.14 electron/(a.u.)$^3$ [23] at 4.2 K as dynamics (SOD) is the same for both sites. Atoms at [0 k 0] have lower electron density. Solid lines represent fit of the SOD to the data above magnetic transition, i.e., in the range 72 – 300 K [31]. Due to the fact that the phonon density of states (DOS) has been neither measured nor calculated for FeAs a Debye approximation was used. Debye temperatures $\theta_D$ are shown in Figure 6. They are almost the same for both models, but they differ between iron sites. However, the difference is not caused by different dynamics, but it is due to the variation of the isomer shift with temperature at various rates for different iron sites. The average Debye temperature amounts to $\langle \theta_D \rangle = 435(5)$ K independently of the model used. The real error might as large as several decades due to the temperature variation of the electron density on the iron nuclei.

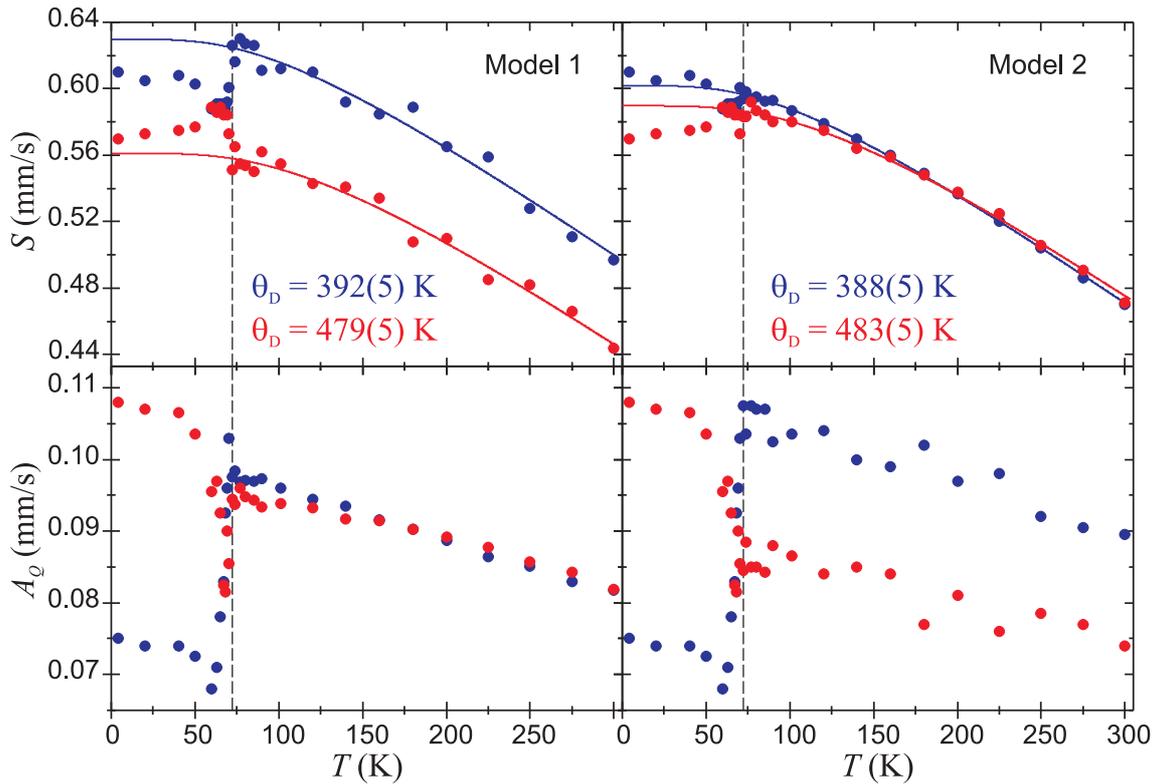

**Figure 6** Spectral shift $S$ and quadrupole coupling constant $A_Q$ are plotted versus temperature for both models. Red color is used for the $[0\,k+\tfrac{1}{2}\,0]$ iron, while blue for the [0 k 0] iron. Dashed vertical lines at 72 K are used to separate magnetically ordered region from the paramagnetic region. Spectral shifts are reported versus room temperature α-Fe. Quadrupole splitting in the non-magnetic region amounts to $\Delta = 6A_Q\sqrt{1+\eta^2/3}$ with $\eta = 0.88(1)$.

Spectral doublets above magnetic transition show intensity anisotropy with the lower velocity line being more intense. This feature was observed in previous Mössbauer works [18, 20]. Due to the large axial anisotropy of EFG the line asymmetry was accounted for fitting the parameter $\text{Re}(g_{1-1})$ as described above. Parameter $\text{Re}(g_{1-1})$ is plotted versus temperature in Figure 7. One can see that it drops to null at the onset of the magnetic transition approached from above and varies with temperature confirming that the line asymmetry has dynamic origin. Upon having solved numerically equation (3) one obtains parameter $\Delta b$ versus temperature. Parameters $\Delta b$ could be used in the harmonic Debye approximation to fit Debye temperatures along the $b$-axis and in the $a$-$c$ plane called here $\theta_b$ and $\theta_{ac}$, respectively. For $\theta_{ac}$ one obtains $\theta_{ac} = 330(1)$ K. Due to the large data scatter the value of $\theta_b$ is not reliable, as it diverges to unphysically large values.

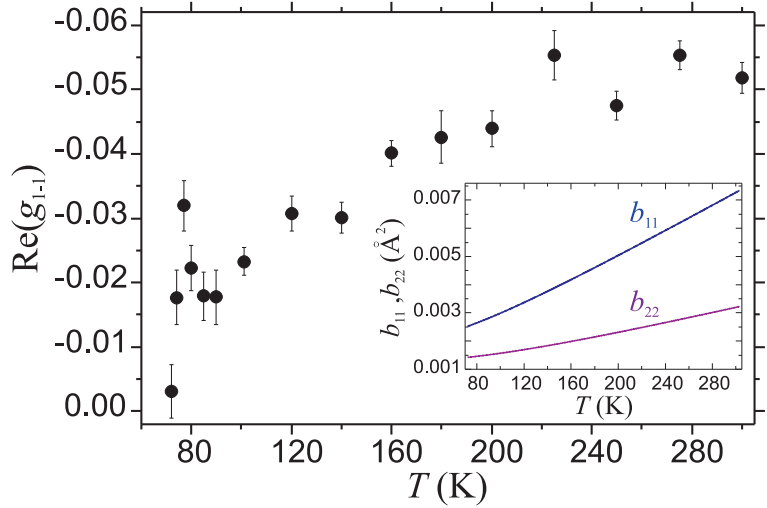

**Figure 7** Plot of the parameter $\text{Re}(g_{1-1})$ versus temperature in the temperature region 72 – 300 K. Inset shows $b_{11}$ and $b_{22}$ versus temperature.

Figure 8 shows $(t\Gamma)/(t_0\Gamma_0) = \langle f \rangle / \langle f_0 \rangle$ versus temperature. Here the parameter $t = N\sigma_0 d_0 (\Gamma_n/\Gamma)\langle f \rangle$ stands for the dimensionless absorber thickness obtained from the transmission integral fits. The symbol $N$ denotes density of the resonant nuclei (in the ground state). The symbol $\sigma_0$ stands for the resonant cross-section, $d_0$ denotes the absorber thickness along the beam and $\Gamma_n$ stands for the natural line-width of 0.097 mm/s. Symbol $\Gamma \geq \Gamma_n$ denotes actual line-width within absorber, while $\langle f \rangle$ stands for the average recoilless fraction, as the absorber has no preferential orientation. The subscript 0 refers to some temperature chosen as the reference temperature. Here the reference temperature is chosen as 72 K. The solid line shows fit within Debye approximation to the points in the range 72 – 300 K (see [31]). One can fit data within harmonic isotropic model with the single Debye temperature $\theta_{iso}$ or within anisotropic harmonic model with separate Debye temperatures $\theta_{ac}$ and $\theta_b$. For the latter case the temperature $\theta_{ac}$ was kept constant at the value obtained fitting parameter $\Delta b$. Both fitted curves are practically identical. The following results were obtained $\theta_{iso} = 368(9)$ K and $\theta_b = 508(68)$ K. Upon having established Debye temperatures $\theta_{ac}$ and $\theta_b$ one can calculate $b_{11}$ and $b_{22}$ versus temperature shown in inset of Figure 7. Corresponding $\langle f_0 \rangle$ amounts to 0.90, while for the isotropic model one obtains $f_0 = 0.87$.

Recoilless fraction increases significantly at the magnetic transition as shown in Figure 8. It means that lattice hardens, while going into magnetically ordered state. Hence, the SOD has to drop to lower values in the magnetically ordered state, but this feature is masked by the simultaneous change of the electron density on the iron nuclei. Simple scaling of either $\langle f_0 \rangle$ or $f_0$ into magnetic region shows that the recoilless fraction in this region approaches unity. One has to observe that the absorber dimensionless thickness and line-width are determined definitely less accurately in the magnetic region. It is worth mentioning that SOD and recoilless fraction Debye temperatures are not identical even for the isotropic case as DOS is differently probed for SOD and recoilless fraction [31]. It seems that $\langle f_0 \rangle$ (and $f_0$) is somewhat overestimated due to application of oversimplified DOS of the Debye approximation. Hence, the parameters $b_{11}$ and $b_{22}$ are underestimated, but the difference $\Delta b$ is more accurate.

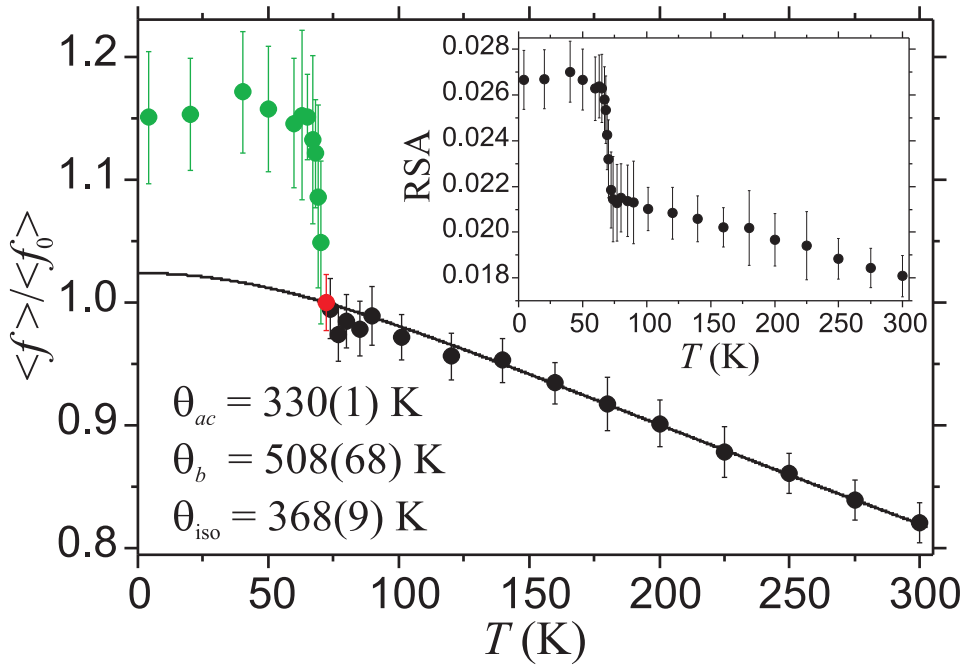

**Figure 8** Plot of the relative recoilless fraction $\langle f \rangle / \langle f_0 \rangle$ versus temperature up to 300 K. Green points correspond to the magnetically ordered region. Red point is the normalization point. Inset shows relative spectral area RSA plotted versus temperature.

Relative spectral area (RSA) is plotted versus temperature as inset of Figure 8. It is defined as:

$$\text{RSA} = \left( \frac{1}{C} \right) \sum_{n=1}^{C} \frac{N_0 - N_n}{N_0}.$$

(4)

Here the symbol $C$ stands for the number of data channels in the folded spectrum. The symbol $N_0$ stands for the baseline i.e. the average number of counts per channel far off the resonance (baseline of the folded spectrum is flat for adopted geometry), while the symbol $N_n$ stands for the number of counts in the channel $n$. All spectra belonging to this set of spectra were recorded with the same source kept under the same conditions, within the same geometry, with the same background under the resonant line (photo-peak was used) and within the linear response range of the detector system. Additional confirmation of the detector system linearity follows from the regular behavior of the $\langle f \rangle / \langle f_0 \rangle$ parameter versus temperature above magnetic transition as shown in Figure 8, i.e., a good reproduction of the Debye model of the lattice vibrations. The velocity scale (and the number of data channels) was the same for all of the above spectra. Hence, one can conclude that the lattice hardening at the magnetic transition is real and does not depend on the particular way of the data treatment as the expression defining parameter RSA does not depend on any physical model. Increase of the parameter RSA could be explained only as the increase of the average recoilless fraction. Estimated detector system non-linearity contributes well below 10 % to the RSA jump at the magnetic transition i.e. much less than the error bar. Note the larger relative jump of RSA in comparison with the corresponding jump of $\langle f \rangle / \langle f_0 \rangle$ at the magnetic transition. It is due to the finite absorber thickness accounted properly by the transmission integral method used to fit the spectra.

Figure 9 shows spectra obtained in the external field of 7 T and for comparison spectra obtained at the same temperature in the null field. Spectrum obtained at 20 K is quite complex indicating rather small susceptibility of the random sample to the external field. Hence, one can conclude that magnetic moments are hard to reorient. It means that spin polarization of the Fe-As bonds is very important here. Spectrum obtained at 100 K has been fitted as described above with non-magnetic hyperfine parameters fixed on values obtained in the null field and the same temperature. The fit is not very good as the field acting on the iron nuclei could be written as $\mathbf{B} = (\mathbf{1} - \boldsymbol{\chi})\mathbf{B}_0$ with the symbol $\mathbf{1}$ denoting unit operator and the symbol $\boldsymbol{\chi}$ standing for the local susceptibility tensor having in principle nine components and being different for two iron sites. The symbol $\mathbf{B}_0$ stands for the applied field. Due to the limited data accuracy the susceptibility was approximated by the scalar replacing tensor $\boldsymbol{\chi}$ by $\boldsymbol{\chi} = \chi \mathbf{1}$. Scalar susceptibilities amount to 0.0214(7) for $[0\,k\,0]$ iron and to 0.0043(7) for $[0\,k+\tfrac{1}{2}\,0]$ iron within Model 1. Model 2 seems to be inappropriate as one gets 0.054(10) and –0.026(10), respectively. Due to the fact that applied field is larger than possible internal field one obtains strong diamagnetic behavior for $[0\,k+\tfrac{1}{2}\,0]$ iron in the Model 2. Such situation has no physical meaning.

Figure 10 shows spectra obtained in the oven. They were taken increasing temperature from 300 K till 1000 K. The last spectrum at the bottom is taken at 300 K as the last one on that velocity scale. Some intermediate spectrum at 1000 K is shown as well as third from the bottom. Sample starts to loose slowly arsenic at about 1000 K and under vacuum (~$10^{-6}$ hPa) transforming into $Fe_2As$ [32]. No metallic iron was found taking spectrum at large velocity scale after the thermal cycle and at 300 K. Total shift, quadrupole coupling constant and relative recoilless fraction (normalized to 72 K) are shown in the whole temperature range in Figure 11. The spectrum taken in the oven at 300 K prior to the thermal cycle was used to obtain the common scale for $\langle f \rangle / \langle f_0 \rangle$. A distinction of two iron sites is impossible above 300 K (since 600 K at least). The Debye temperature for SOD obtained by fit to all data points above and at 72 K amounts to 428(3) K. Hence, it is very close to the previously obtained value $\langle \theta_D \rangle = 435(5)$ K with data points limited to 300 K. Therefore, the variation of the electron density versus temperature on the iron nuclei does not change significantly above room temperature. The quadrupole coupling constant shows some curvature at high temperatures while approaching some small value at some limiting temperature. It is interesting to note, that the parameter $\Delta b$ seems to approach constant value at high temperature (see inset of Figure 11). It means, that strict harmonic approximation does not apply above 300 K and one has to adopt temperature dependent Debye temperature in this temperature region i.e. *quasi*-harmonic approximation. This is an indication that DOS starts to evolve at higher temperatures. It is likely that the Debye temperature $\theta_b$ drops at higher temperatures leading to the observed effect. On the other hand, the whole set of relative recoilless fraction data since 72 K could be fitted nicely by single isotropic Debye temperature even including dubious points at 1000 K . Debye temperature obtained this way is almost the same as obtained previously for the data set limited to the upper temperature of 300 K.

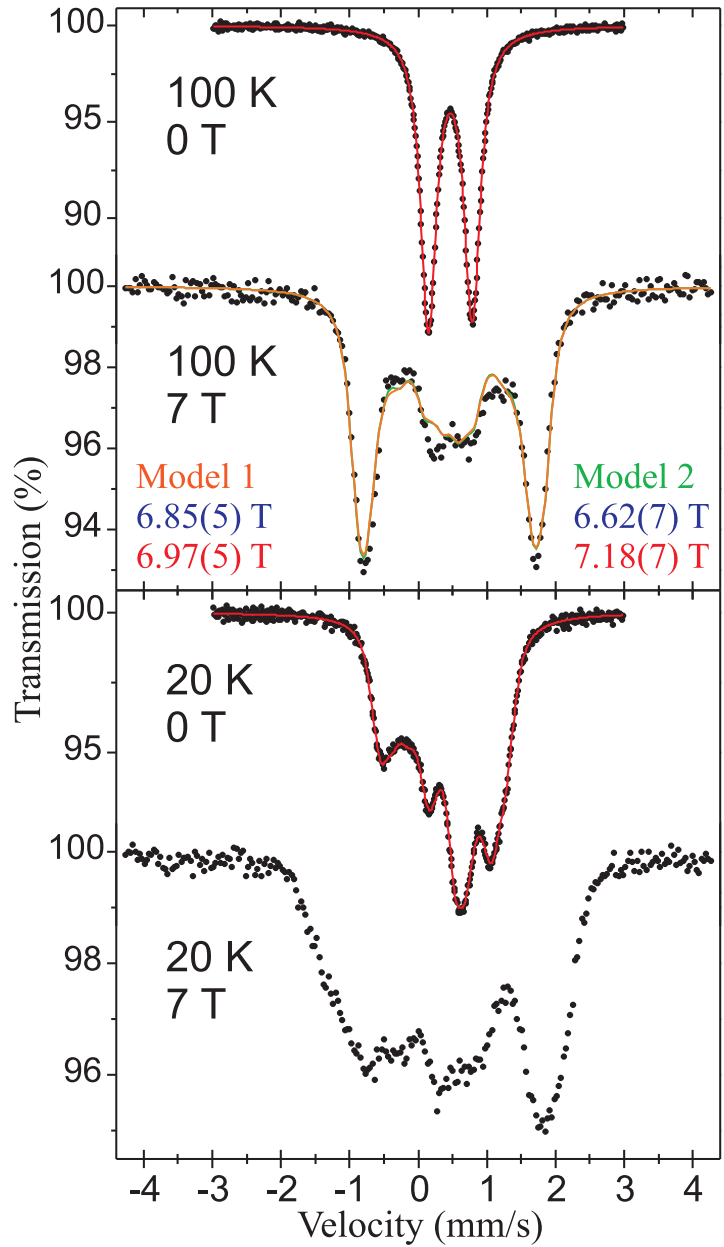

**Figure 9** Mössbauer spectra of FeAs obtained in the external field 7 T oriented anti-parallel to the beam propagation direction. For comparison corresponding spectra obtained at the same temperature, albeit in the null field are shown. The absorber for measurements in the external field was thinner than the one used in the null field. Red color refers to value of the magnetic field on $[0\,k+\tfrac{1}{2}\,0]$ iron nucleus and blue to $[0\,k\,0]$ iron nucleus.

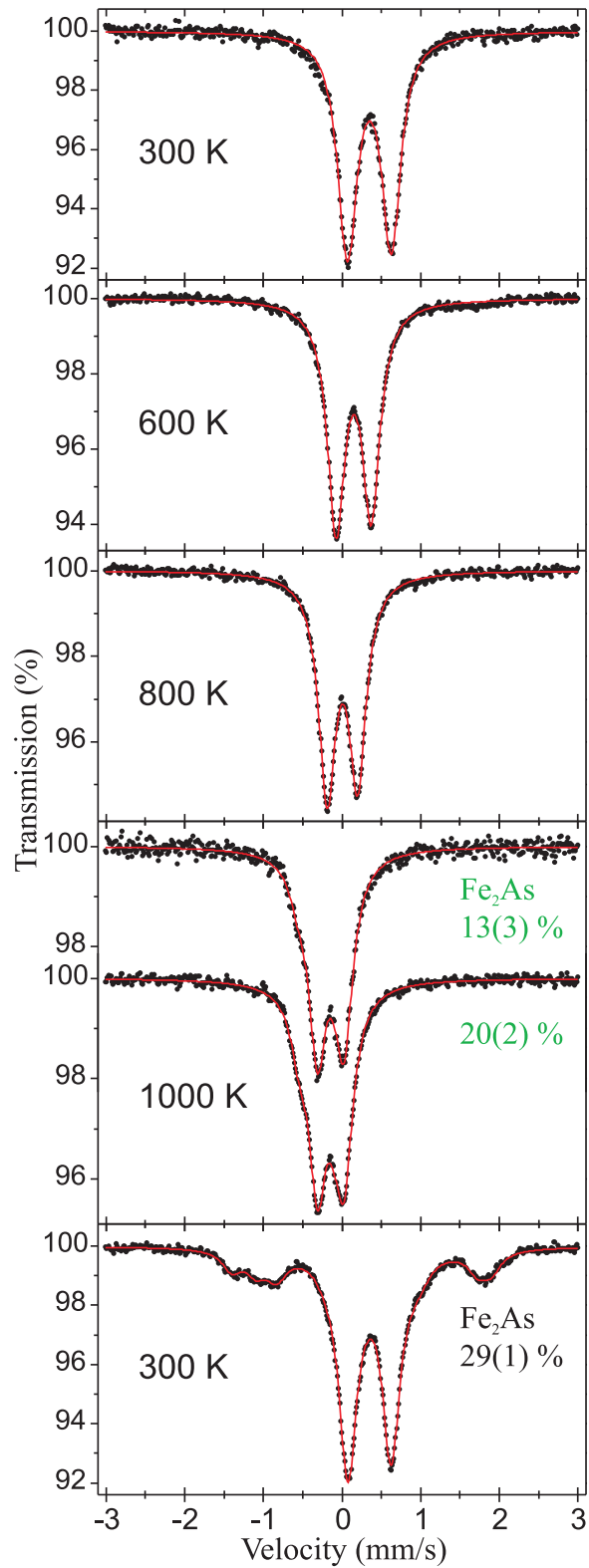

**Figure 10** Mössbauer spectra obtained in oven. The lowest spectrum has been collected after measuring remaining spectra in sequence of the increasing temperature. Approximate contribution of the Fe$_2$As is indicated for spectra of 1000 K and the last 300 K spectrum. The contribution (especially at high temperature) is subject to differences in the average recoilless fraction. Note that Fe$_2$As steadily accumulated during acquisition of spectra at 1000 K. Upper spectrum at 1000 K comes from about half of the total accumulation time at this temperature amounting to 48 h.

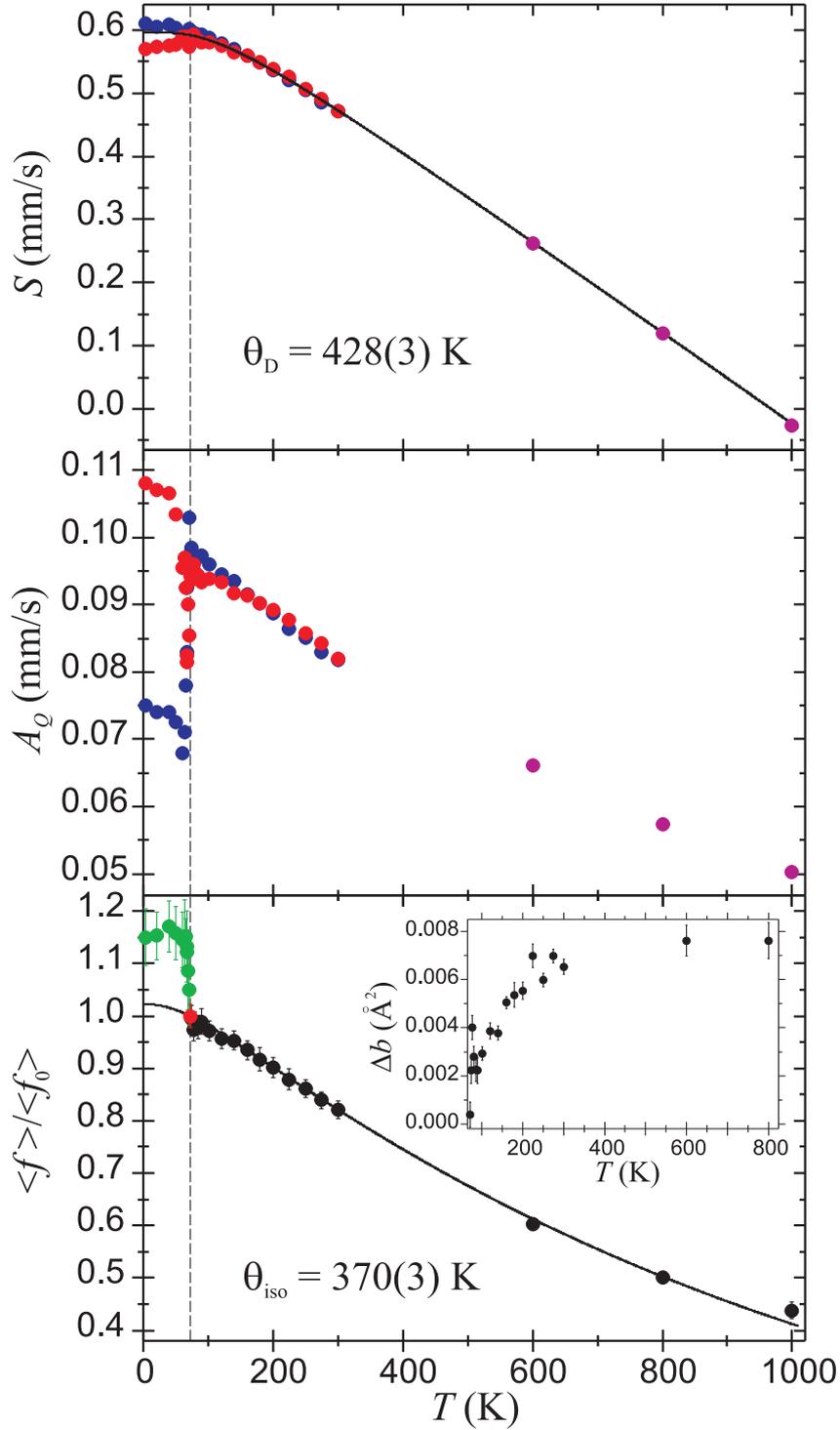

**Figure 11** Spectral shift, quadrupole coupling constant $A_Q$ and $\langle f \rangle / \langle f_0 \rangle$ are plotted versus temperature up to 1000 K. Red color is used for the $[0\,k+\tfrac{1}{2}\,0]$ iron, while blue for the $[0\,k\,0]$ iron. Dashed vertical lines are used to separate magnetically ordered region from the paramagnetic region. Pink color is used in the high temperature region, where two iron sites are indistinguishable. For the lowest part showing $\langle f \rangle / \langle f_0 \rangle$ red point indicates normalization point (72 K), while green points belong to the magnetically ordered region. The point 300 K (in the oven) was used to match both parts (up to 300 K and above) for two different absorbers. Inset shows $\Delta b$ versus temperature ranging from 72 K till 800 K.

## 5. Conclusions

The most striking result is shown in Figure 4. The iron hyperfine field along the electronic spin spiral varies enormously in amplitude in the magnetically ordered region. The pattern resembles symmetry of 3d electrons in the *a-b* plane with the significant distortion caused by the arsenic bonding p electrons. Influence of the spin polarized conduction band is particularly important for crystallographically perturbed $[0\,k+\frac{1}{2}\,0]$ iron sites, where patterns are much smoother than for $[0\,k\,0]$ iron sites. These two sites differ by the electron density on the iron nucleus at low temperatures with the $[0\,k\,0]$ site having lower density. Häggström *et al.* [20] anticipated complex distribution and variation of the hyperfine fields along the spiral i.e. variation of the hyperfine field with the changing orientation of the local magnetic moment. Hence, one can conclude that the magnetic hyperfine constant $A(\phi)$ transforming electronic magnetic moment into the hyperfine field on the iron nucleus is strongly direction dependent and follows geometry of the magnetically polarizable bonds. It is probably not accidental that the highest order of the tensors describing anisotropy of the hyperfine field amounts to $L=4$. Wave functions of the bonding electrons are linear combinations of the s-p-d electrons. Hence, the fourth order terms in the angular space are made as the highest order terms upon having squared above wave functions. It seems that local susceptibility on the iron nuclei remains anisotropic even at 100 K i.e. well above transition to the paramagnetic state.

Another unusual feature is strong coupling between magnetism and lattice dynamics i.e. very strong phonon-magnon interaction. Iron is coupled in the isotropic fashion to the lattice in the magnetic region. The coupling softens mainly in the *a-c* plane upon transition to the paramagnetic state, anisotropy follows roughly harmonic behavior till about 300 K and saturates at high temperatures (constant $\Delta b$) indicating *quasi*-harmonic behavior at high temperatures. No true anharmonic forces were found. It is interesting to note that this lattice hardening does not show clearly in the specific heat [8]. Probably it is masked in some way by magnetic contributions. Lattice hardening occurs without visible structural transition. Strong phonon-magnon coupling is obviously the feature of the iron arsenic bonds playing important role in the large class of the iron-based superconductors [21].

Static critical exponents suggest some underlying transition leading to the magnetic order. Due to the lack of the structural changes one can envisage some subtle order-disorder transition with very small latent heat and hysteresis driven by the itinerant charge/spin ordering.

The sample starts to loose arsenic at about 1000 K under vacuum, what might be explanation for the specific heat anomaly observed at high temperature [8].

It seems that phonon DOS has to be measured at various temperatures. Simultaneously measurements of the similar DOS for magnons in the magnetically ordered state are needed. Such data could shed more light on the role of the phonon-magnon coupling and role of the orbital terms in the lattice hardening – magnetostriction. Precise structure determination versus temperature is needed to resolve the problem of departure from the exact *Pnma* symmetry. Arsenic nuclear magnetic resonance might provide information about the transferred hyperfine magnetic fields and electric quadrupole interaction on arsenic atoms as the sole stable isotope of arsenic $^{75}$As has nuclear spin of 3/2 and quite large spectroscopic quadrupole moment. It would be interesting to know how the large and anisotropic magnetic polarization of the Fe-As bonds transforms with the varying structure and band filling into system being able to create Cooper pairs i.e. into superconductors.

**Acknowledgments**

This project was financially supported by the National Science Center of Poland under the Grant No. DEC-2011/03/B/ST3/00446. The superconducting magnet 7TL-SOM2-12 MOSS was purchased thanks to financial support by the European Regional Development Fund under the Infrastructure and Environment Programme. Development of the MsAa-4 spectrometer was partly financed by the Polish Ministry of Science and Higher Education under the Grant No. R15-002-03.